\documentclass{elsart}

\usepackage[square,comma]{natbib}
\usepackage{graphicx}
\usepackage{pxfonts}
\usepackage{lineno}

\usepackage{amssymb}

\journal{}

\begin{document}
\thispagestyle{empty}
\begin{Large}
\textbf{DEUTSCHES ELEKTRONEN-SYNCHROTRON}

\textbf{\large{Ein Forschungszentrum der Helmholtz-Gemeinschaft}\\}
\end{Large}

DESY 11-224

November 2011

\begin{eqnarray}
\nonumber &&\cr \nonumber && \cr \nonumber &&\cr
\end{eqnarray}
\begin{eqnarray}
\nonumber
\end{eqnarray}
\begin{center}
\begin{Large}
\textbf{Extension of self-seeding to hard X-rays} $\mathbf{> 10}$
\textbf{keV as a way to increase user access at the European XFEL}
\end{Large}
\begin{eqnarray}
\nonumber &&\cr \nonumber && \cr
\end{eqnarray}

\begin{large}
Gianluca Geloni,
\end{large}
\textsl{\\European XFEL GmbH, Hamburg}
\begin{large}

Vitali Kocharyan and Evgeni Saldin
\end{large}
\textsl{\\Deutsches Elektronen-Synchrotron DESY, Hamburg}
\begin{eqnarray}
\nonumber
\end{eqnarray}
\begin{eqnarray}
\nonumber
\end{eqnarray}
ISSN 0418-9833
\begin{eqnarray}
\nonumber
\end{eqnarray}
\begin{large}
\textbf{NOTKESTRASSE 85 - 22607 HAMBURG}
\end{large}
\end{center}
\clearpage
\newpage

\begin{frontmatter}




\title{Extension of self-seeding to hard X-rays $\mathbf{> 10}$ keV as a way to increase user access at the European XFEL}


\author[XFEL]{Gianluca Geloni\thanksref{corr},}
\thanks[corr]{Corresponding Author. E-mail address: gianluca.geloni@xfel.eu}
\author[DESY]{Vitali Kocharyan}
\author[DESY]{and Evgeni Saldin}

\address[XFEL]{European XFEL GmbH, Hamburg, Germany}
\address[DESY]{Deutsches Elektronen-Synchrotron (DESY), Hamburg,
Germany}

\begin{abstract}
We propose to use the self-seeding scheme with single crystal
monochromator at the European X-ray FEL to produce monochromatic,
high-power radiation at $16$ keV. Based on start to end simulations
we show that the FEL power of the transform-limited pulses can reach
about $100$ GW by exploiting tapering in the tunable-gap baseline
undulator. The combination of high photon energy, high peak power,
and very narrow bandwidth opens a vast new range of applications,
and includes the possibility to considerably increase the user
capacity and fully exploit the high repetition rate of the European
XFEL. In fact, dealing with monochromatic hard X-ray radiation one
may use crystals as deflectors with minimum beam loss. To this end,
a photon beam distribution system based on the use of crystals in
the Bragg reflection geometry is proposed for future study and
possible extension of the baseline facility. They can be repeated a
number of times to form an almost complete (one meter scale) ring
with an angle of $20$ degrees between two neighboring lines. The
reflectivity of crystal deflectors can be switched fast enough by
flipping the crystals with piezo-electric devices similar to those
for X-ray phase retarders at synchrotron radiation facilities. It is
then possible to distribute monochromatic hard X-rays among $10$
independent instruments, thereby enabling $10$ users to work in
parallel. The unmatched repetition rate of the European XFEL would
be therefore fully exploited.
\end{abstract}

%
%
%
\end{frontmatter}



\section{\label{sec:intro} Introduction}

\begin{figure}[tb]
\includegraphics[width=1.0\textwidth]{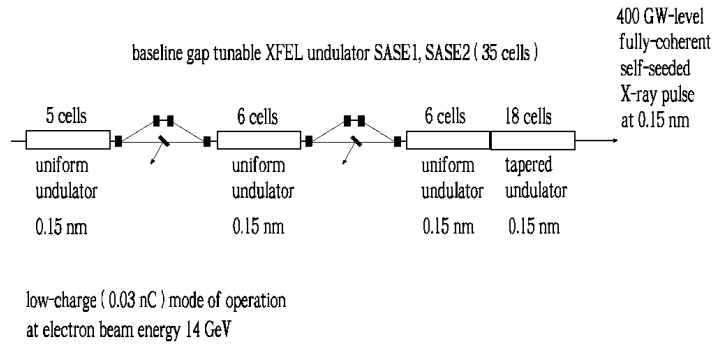}
\caption{Sketch of an undulator system for high power mode of
operation at a photon energy of $8$ keV. The method exploits a
combination of cascade self-seeding scheme with single crystal
monochromator and undulator tapering technique. The amplification-
monochromatization cascade scheme is distinguished, in performance,
by spectral purity of the output radiation and smaller heat loading
of crystals. } \label{F1}
\end{figure}

Radiation from SASE XFEL consists of many independent spikes in both
the temporal and spectral domains. Self-seeding is a promising
approach to significantly narrow the SASE bandwidth to produce
nearly transform-limited pulses \cite{SELF}-\cite{FENG}. We
discussed the implementation of a single-crystal self-seeding scheme
in the hard X-ray lines of European XFEL in \cite{OURY3,OURY5}. For
this facility, transform-limited pulses are particularly valuable,
since they naturally support the extraction of more FEL power than
at saturation by exploiting tapering in the tunable-gap baseline
undulators \cite{TAP1}-\cite{WANG}. Tapering is implemented as a
stepwise change of the undulator gap from segment to segment.
Simulation results presented in \cite{OURY3,OURY5} show that the FEL
power of the transform-limited X-ray pulses may be increased up to
$0.4$ TW by operating with the tapered baseline undulator SASE1 (or
SASE2). In particular, it is possible to create a source capable of
delivering fully-coherent, $7$ fs (FWHM)-long X-ray pulses with
$2\cdot 10^{12}$ photons per pulse at a wavelength of $0.15$ nm,
Fig. \ref{F1}.

\begin{figure}[tb]
\includegraphics[width=1.0\textwidth]{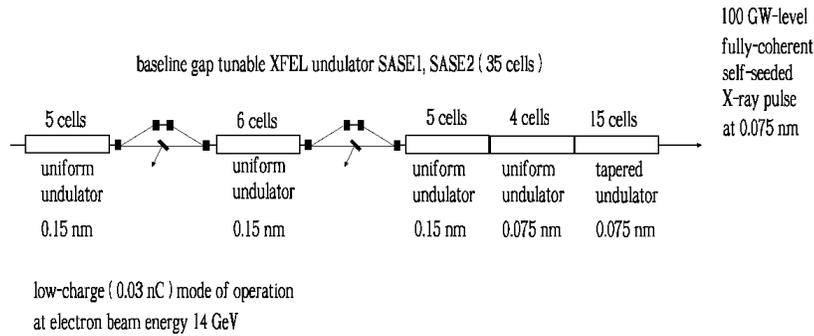}
\caption{Sketch of an undulator system for generating high power,
highly monochromatic hard X-ray beam at a photon energy of $16$
keV.} \label{F2}
\end{figure}

We can apply the same scheme to harder X-rays, and obtain $100$ GW
fully-coherent X-ray pulses at a wavelength of $0.075$ nm.  In this
paper we propose to perform monochromatization at $0.15$ nm with the
help of self-seeding, and amplify the seed in a first part of the
output undulator. The amplification process can be stopped at some
position well before the FEL reaches saturation, where the electron
beam gets considerable bunching at the $2$nd harmonic of the
coherent radiation. A second part of the output undulator tuned to
the $2$nd harmonic frequency, follows beginning at that position,
and is used to obtain $2$nd harmonic radiation at saturation. One
can prolong the exchange of energy to the advantage of the photon
beam by tapering the last part of the output undulator on a segment
by segment basis. Fig. \ref{F2} shows the design principle of our
self-seeding setup for harder photon energy mode of operation.  Two
self-seeding cascades, identical to those considered in
\cite{OURY3,OURY5} (see Fig. \ref{F1}), are followed by the same
output undulator with changed gap configuration, compared to Fig.
\ref{F1}.

An advantage of the proposed scheme is the possibility to increase
user capacity. In this paper we describe a photon beam distribution
system, which may allow to switch the hard X-ray beam quickly among
many experiments in order to make a more effective use of the
facility. Monochromaticity is the key for implementing multi-user
operation in the hard X-ray range, which can be granted by using
crystal deflectors and small absorption of the radiation in crystals
at photon energies larger than $15$ keV. In contrast to the
broadband SASE bandwidth, transform-limited  hard X-ray bandwidths
are of order of $0.007 \%$, and match the Bragg width of crystals.
Thus, using $0.05$ mm thick diamond crystals one may obtain two
beams, one transmitted and one Bragg reflected, with minimum
intensity loss. We suggest to flip crystals for switching
reflectivity similarly as for polarization switching techniques with
X-ray phase retarders at synchrotron radiation facilities, that is
based on the use of piezo-electric components. Crystal deflectors
can be repeated a number of times to form an almost complete ring.
It is then possible to distribute monochromatic hard X-rays among
ten independent experiments.

\section{\label{sec:due} Scheme for multiplying the hard X-ray beams to serve more users simultaneously}

In this section we describe a concept for a photon beam distribution
system, which may allow to switch the FEL beam quickly among many
instruments in order to make a more effective use of the facility.
The high photon energy, and monochromaticity  of the output
radiation are the key for reaching such result.

\begin{figure}[tb]
\includegraphics[width=1.0\textwidth]{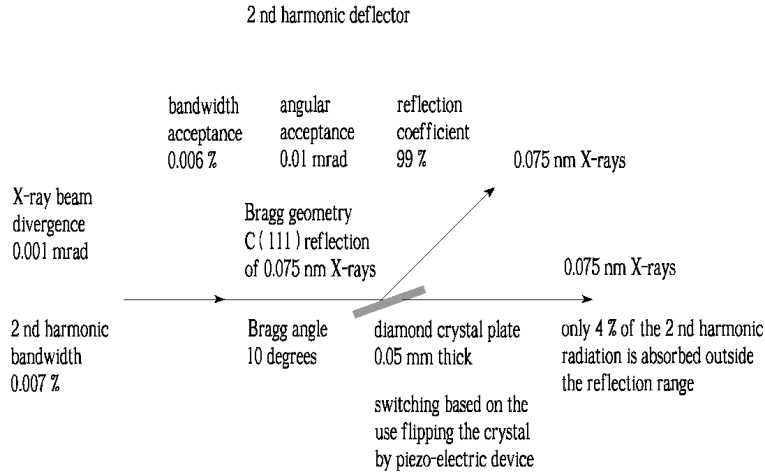}
\caption{Concept for a hard photon beam deflector based on the use
of a crystal in Bragg reflection geometry.  High energy (16 keV)
X-rays are associated with large penetration power, which makes it
possible to use relatively thick diamond deflectors. The thickness
of one crystal is $0.05$ mm. The deflector can be simply switched
off by tilting it, which results in a change of the angle of
incidence of the radiation. The angular change necessary for
switching  on and off the reflected beam is less than $0.1$ mrad.
Only $4 \%$ of incoming $16$ keV radiation is absorbed outside the
reflection range. } \label{F3}
\end{figure}
In fact, dealing with $16$ keV monochromatic radiation one may use
crystals as deflectors, Fig. \ref{F3}. The deflector is constituted
by a diamond plate with a thickness of $0.05$ mm. The crystal is
used in Bragg reflection geometry and exploits the C(111) reflection
plane. For the C(111) reflection, the angular acceptance of the
crystal deflector is of the order of $10~\mu$rad, and the spectral
bandwidth is about $0.006 \%$.  As a result, the angular acceptance
of the deflector is much wider compared to the photon beam
divergence, which is of the order of a microrad. The bandwidth of
the hard X-ray pulse is of order of $0.007 \%$ and matches the Bragg
width of the crystal. In this case, more than $99 \%$ of the peak
reflectivity can be achieved and only $4 \%$ of the incoming $16$
keV radiation is absorbed outside of the reflection range. Thus, by
use of a $0.05$ mm-thick diamond crystal one may obtain two beams, a
transmitted as well as a Bragg reflected beam, with minimal
intensity loss.

One of the biggest advantages of using a crystal deflector in Bragg
geometry is that the reflected beam can be switched on and off by
changing the crystal angle only. The typical angular change
necessary for the switching is less than $0.1$ mrad. This opens a
new possibility of fast switching of the reflectivity, which is
necessary for many-user operation. In order to achieve a stable
photon beam deflection, the rotation error must be less than $0.01$
mrad. Existing technology enables rotating crystals to satisfy this
requirements. For example, at synchrotron radiation facilities,
X-ray phase retarder crystals are driven by piezo-electric devices
operated at hundred $Hz$ repetition rate, which flip the crystals
with a rotation error of about a fraction of a micro-radian
\cite{HIRA}.

\begin{figure}[tb]
\includegraphics[width=1.0\textwidth]{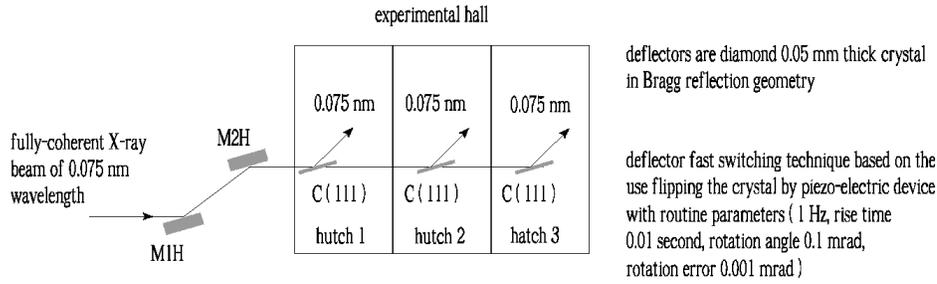}
\caption{The proposed hard X-ray SASE1 (or SASE2) undulator
beamline, consisting of a series of three or more crystals in Bragg
reflection geometry. A photon beam distribution system based on
flipping crystals can provide an efficient way to obtain a many-user
facility. The monochromaticity of the output at $16$ keV constitutes
the key for reaching such result.} \label{F4}
\end{figure}
A photon beam distribution system based on flipping crystals can
provide an efficient way to obtain a many-user facility. A possible
layout is shown in Fig. \ref{F4}. The output radiation passes
through the distribution system, consisting of a series of crystals
in Bragg geometry.  Photon macro-pulses at $16$ keV photon energy
can then be fed into $10$ separate beamlines. The switching crystals
need to flip at frequency $1$ Hz, so that each user receives one
macropulse per second.  It should be noted that the single crystal
provides a sufficiently large deflection angle (of order of $20$
degrees), so that the problem of separation of neighboring beamlines
does not exist.

\begin{figure}[tb]
\includegraphics[width=1.0\textwidth]{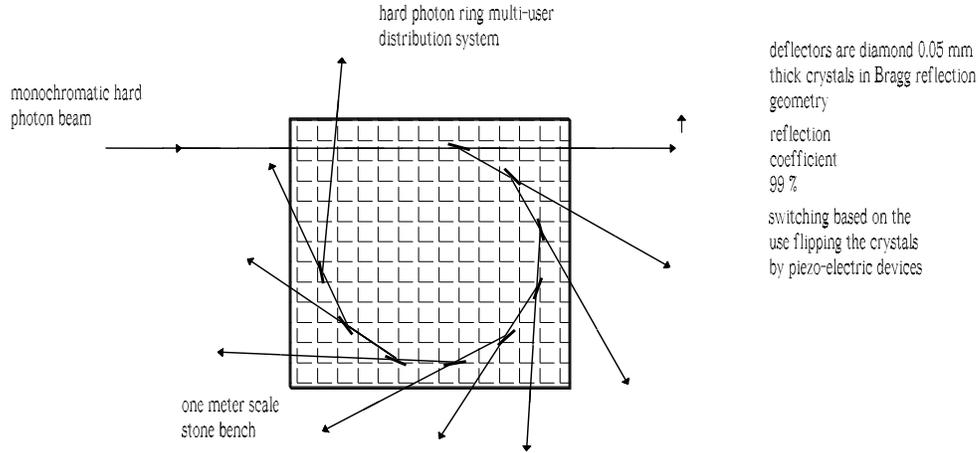}
\caption{Top view of a hard-photon "ring" distribution system.
Separation between two neighboring lines is about $20$ degrees. The
layout of the user instruments follows a similar approach as for
synchrotron radiation sources. } \label{F5}
\end{figure}
A second possible layout of a future hard X-ray laboratory based on
a hard-photon "ring" distribution system is shown in Fig. \ref{F5}.
One can use a number of crystal reflectors to form an almost
complete ring. The photon beam transport line guiding photons from
SASE1 (or SASE2) to the experimental hall is connected tangentially
to one of the straight section of the photon ring, and the beam is
injected by the reflecting crystal. Using flipping crystals in each
photon ring cell it is possible to quickly switch the photon beam
from one instrument to the other, thus providing many-user
capability. This layout of the laboratory follows a similar approach
as for synchrotron light sources.

Finally, it should be remarked that the proposed beam distribution
system operates at fixed wavelength. However as reported in
\cite{LCLS2}: "for many scattering experiments it is not necessary
to continuously vary the X-ray wavelength or fine-tune to a core
shell resonance. Generally, however it is desirable to have a large
photon energy exceeding $10$ keV. This optimizes probing condensed
matter systems on the atomic length scale by minimizing deleterious
absorption, while preserving scattering cross sections". Obviously,
our idea to increase the user access by means of a "photon ring" has
advantages for such kind of experiments and can be applied to the
European XFEL as well as to the LCLS-II design.

\section{\label{sec:tre} Feasibility study for a self-seeding setup at photon energy 16
keV}

In this Section we report on a feasibility study performed with the
help of the FEL code GENESIS 1.3 \cite{GENE} running on a parallel
machine. We will present a feasibility study for a short-pulse mode
of operation of the SASE1 and SASE2 FEL lines of the European XFEL,
based on a statistical analysis consisting of $100$ runs. The
overall beam parameters used in the simulations are presented in
Table \ref{tt1}. We refer to the setup in Fig. \ref{F2}. Up to the
output undulator, simulations are identical to those already
presented in \cite{OURY5}. The choice of the FODO lattice parameters
is also kept identical. This is in agreement with the present
concept to use the self-seeding setup at the European XFEL in
\cite{OURY5} to produce monochromatic beams using harmonic
generation simply by acting on the gap of the output undulator.

\begin{table}
\caption{Parameters for the low-charge mode of operation at the
European XFEL used in this paper.}

\begin{small}\begin{tabular}{ l c c}
\hline & ~ Units &  ~ \\ \hline
Undulator period      & mm                  & 40     \\
Periods per cell      & -                   & 125   \\
K parameter (rms)     & -                   & 2.15  \\
Total number of cells & -                   & 35    \\
Intersection length   & m                   & 1.1   \\
Wavelength            & nm                  & 0.15  \\
Energy                & GeV                 & 14.0 \\
Charge                & pC                  & 28\\
\hline
\end{tabular}\end{small}
\label{tt1}
\end{table}

\begin{figure}[tb]
\includegraphics[width=0.5\textwidth]{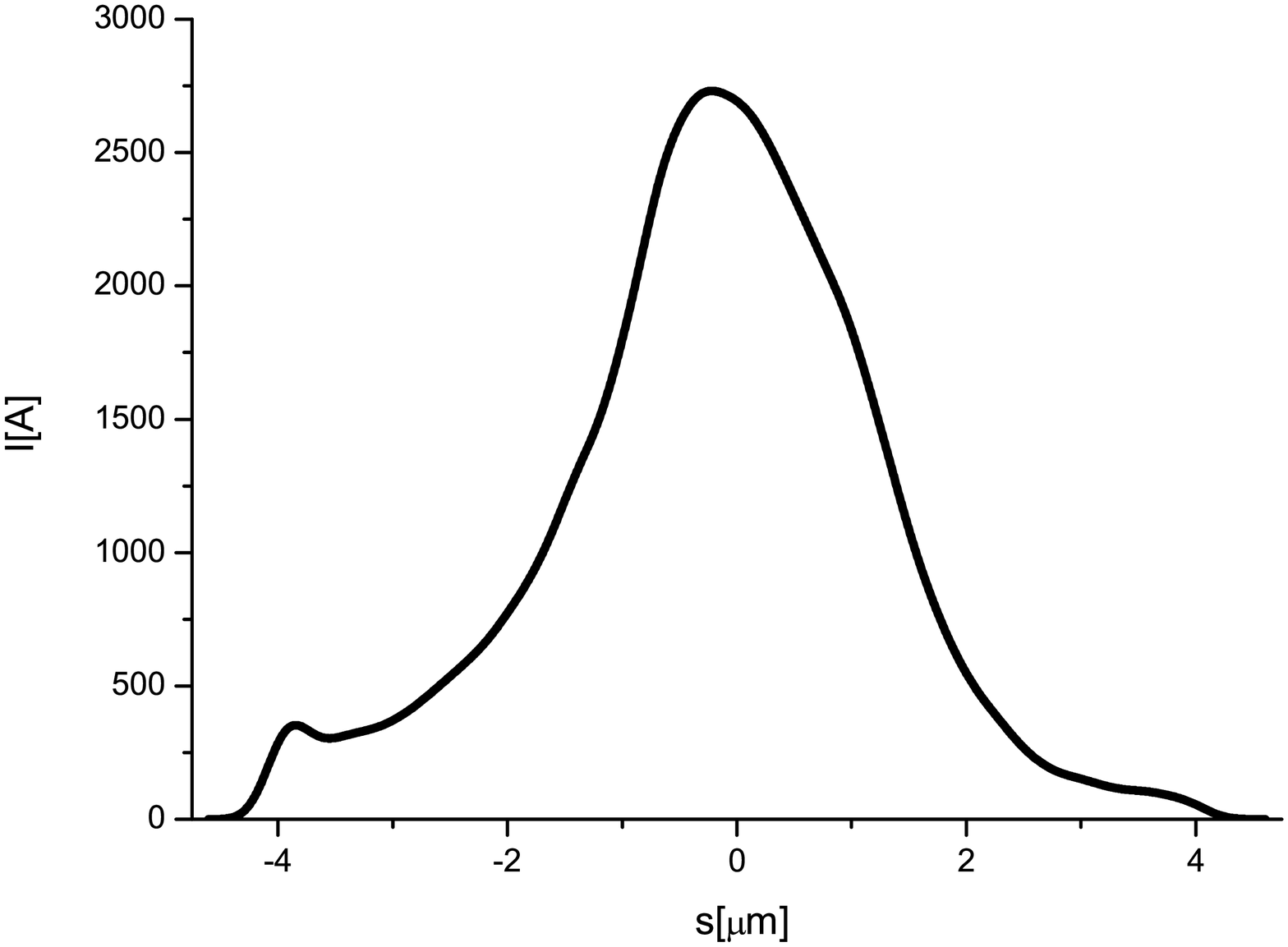}
\includegraphics[width=0.5\textwidth]{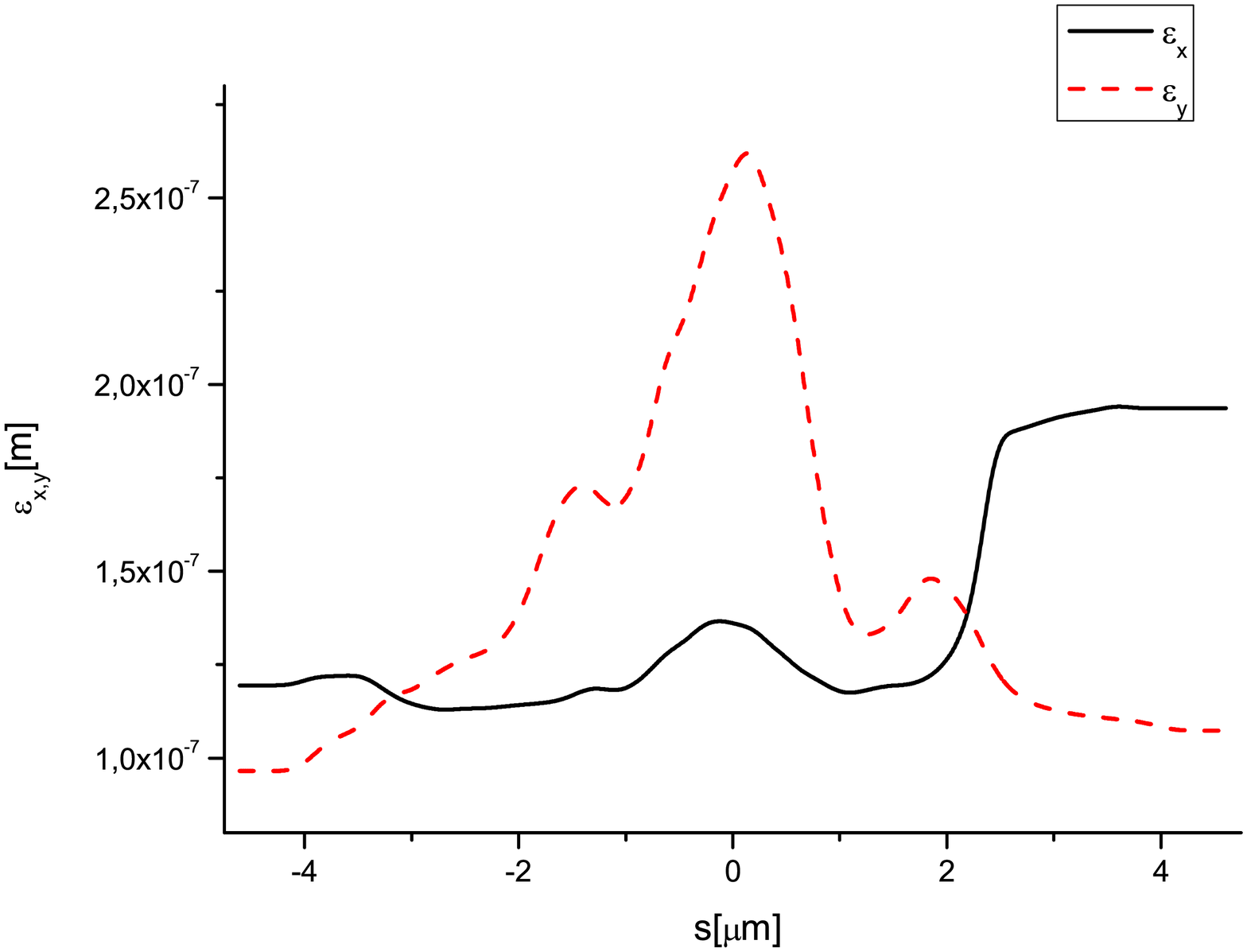}
\includegraphics[width=0.5\textwidth]{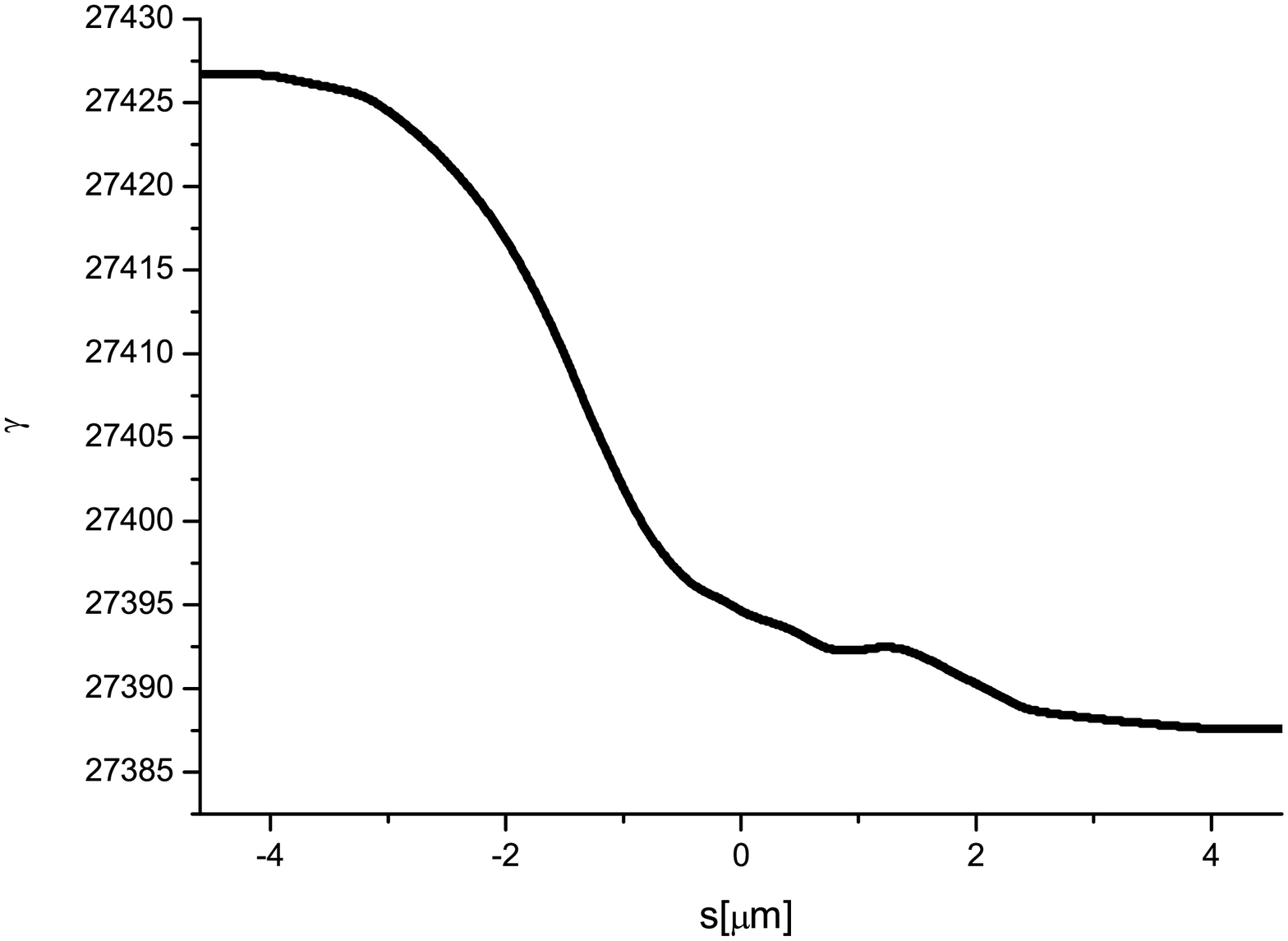}
\includegraphics[width=0.5\textwidth]{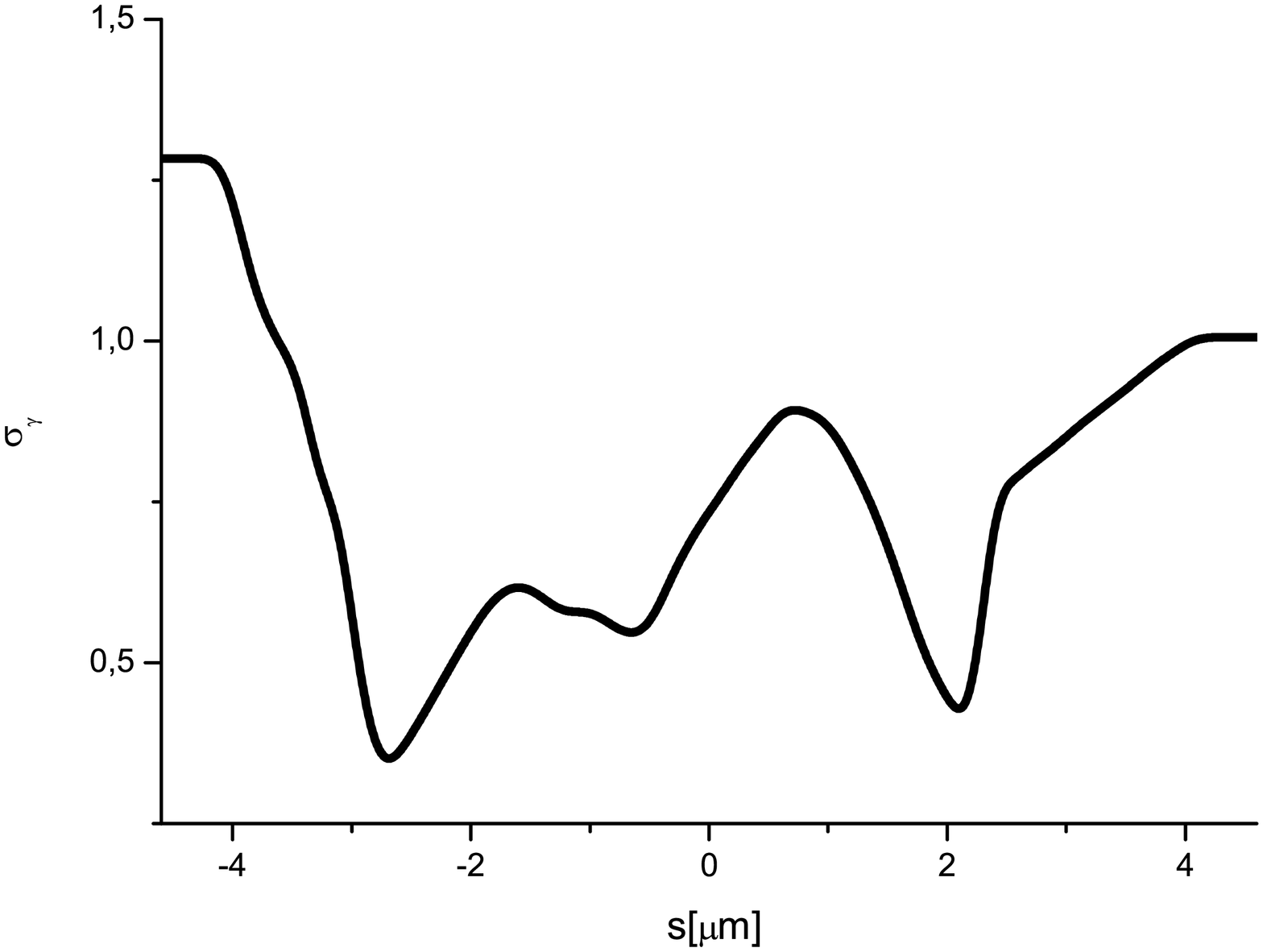}
\caption{Results from electron beam start-to-end simulations at the
entrance of SASE1 and SASE2 \cite{S2ER}. (Top Left) Current profile.
(Top Right) Normalized emittance as a function of the position
inside the electron beam. (Bottom Left) Energy profile along the
beam. (Bottom right) Electron beam energy spread profile.}
\label{s2E}
\end{figure}
\begin{figure}[tb]
\includegraphics[width=1.0\textwidth]{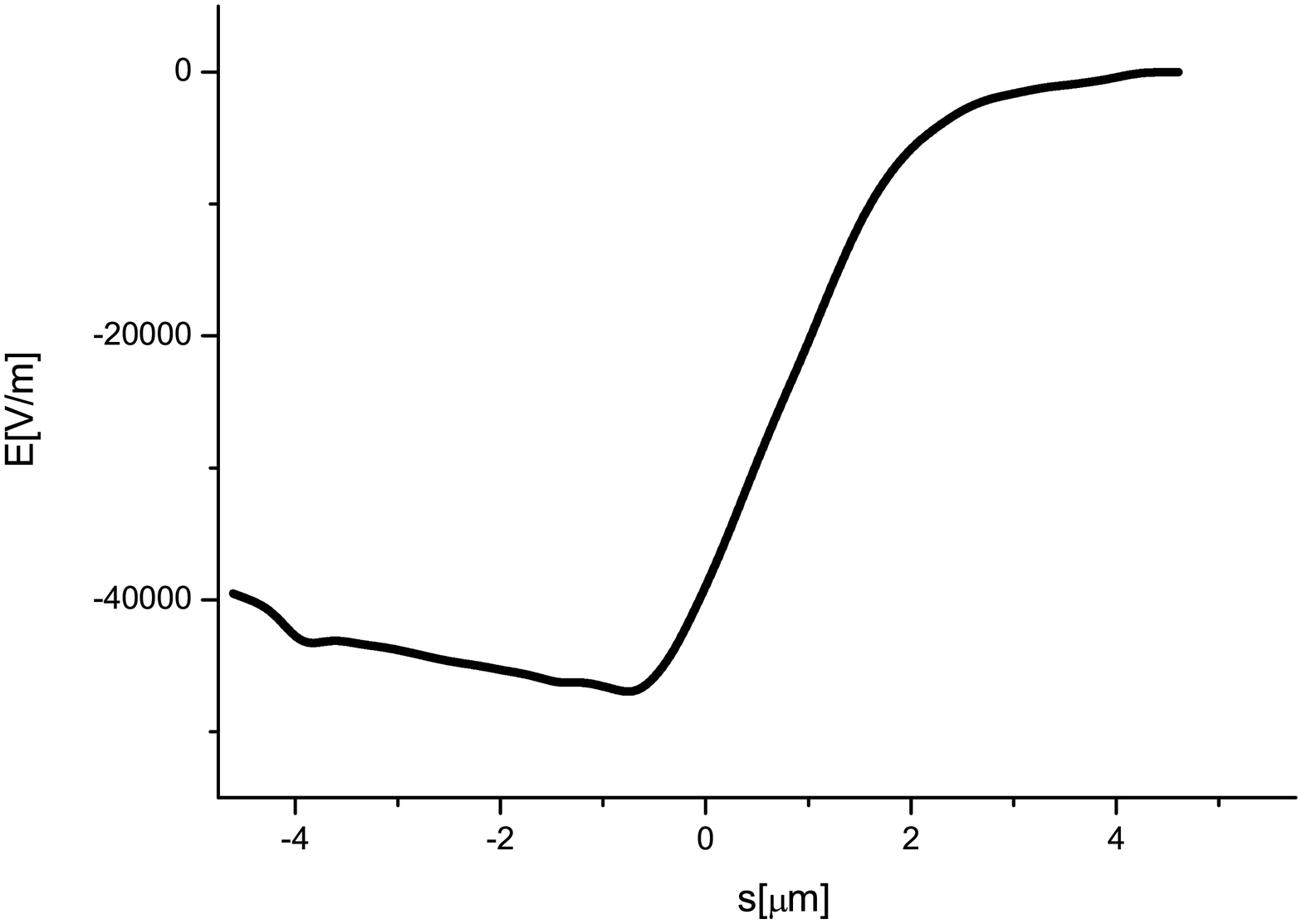}
\caption{Resistive wakefields in the SASE 1 ( SASE 2) undulator
\cite{S2ER}} \label{wake}
\end{figure}
The expected beam parameters at the entrance of the SASE1 and SASE2
undulators are shown in Fig. \ref{s2E}, \cite{S2ER}. Wakes inside
the undulators are also accounted for and expected to obey the
dependence in Fig. \ref{wake}, \cite{S2ER}.

\begin{figure}[tb]
\includegraphics[width=0.5\textwidth]{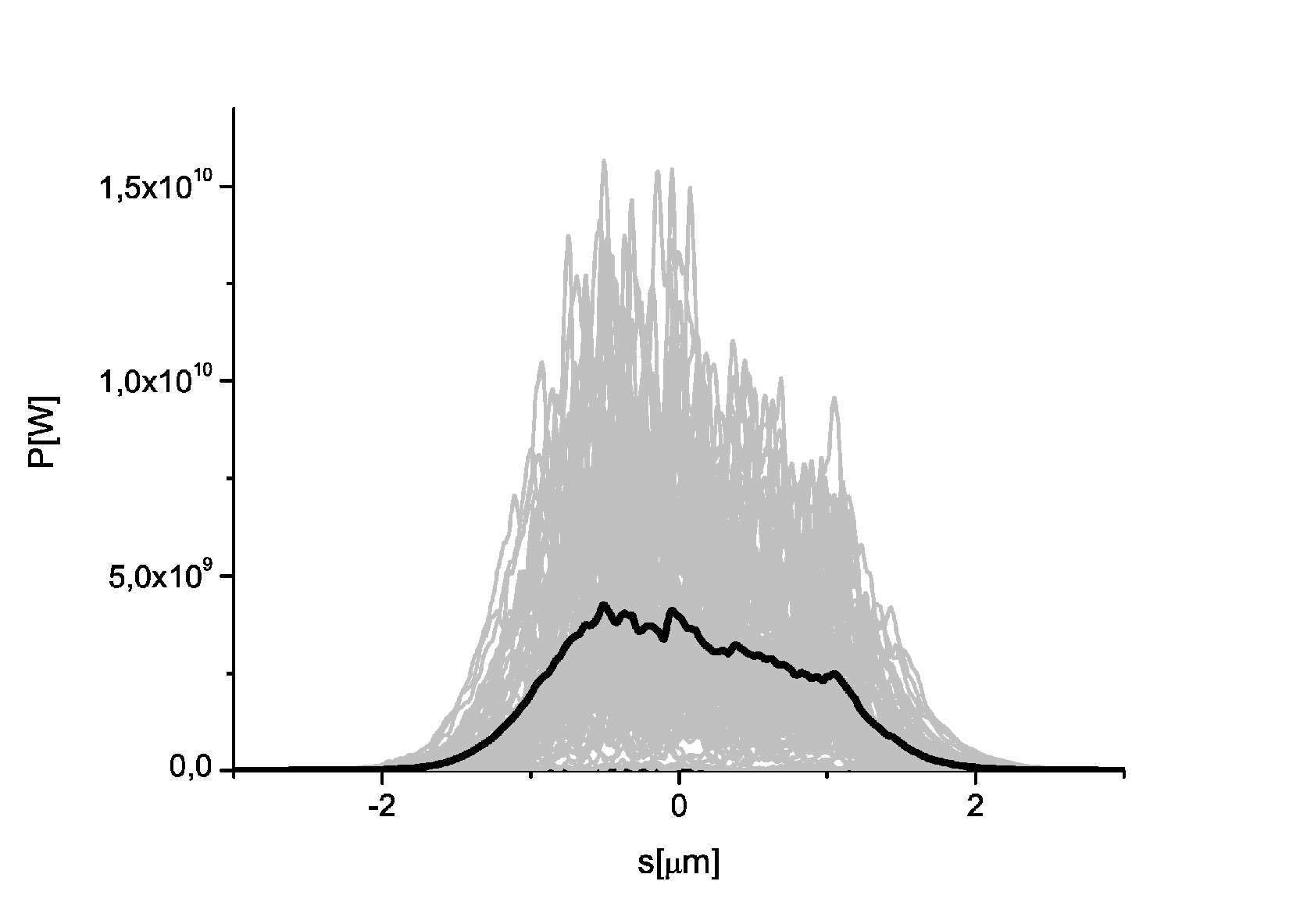}
\includegraphics[width=0.5\textwidth]{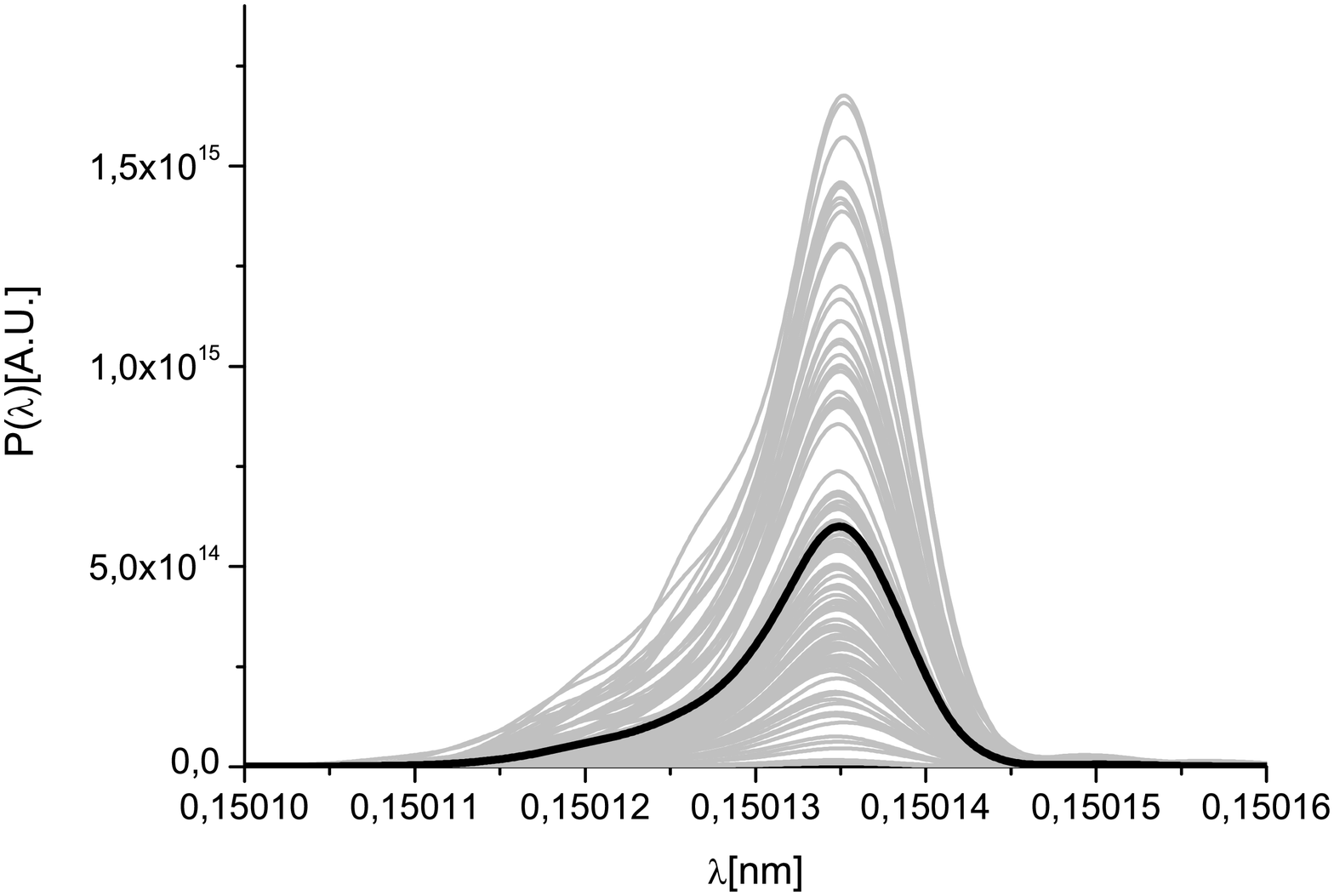}
\caption{(Left plot) Output power and (right plot) output spectrum
after the first part of the output undulator ($5$ cells) tuned at
$0.15$ nm. Grey lines refer to single shot realizations, the black
line refers to the average over a hundred realizations.}
\label{Out1}
\end{figure}
The output power and spectrum after the first part of the output
undulator ($5$ cells) tuned at $0.15$ nm is shown in Fig.
\ref{Out1}.

\begin{figure}[tb]
\includegraphics[width=1.0\textwidth]{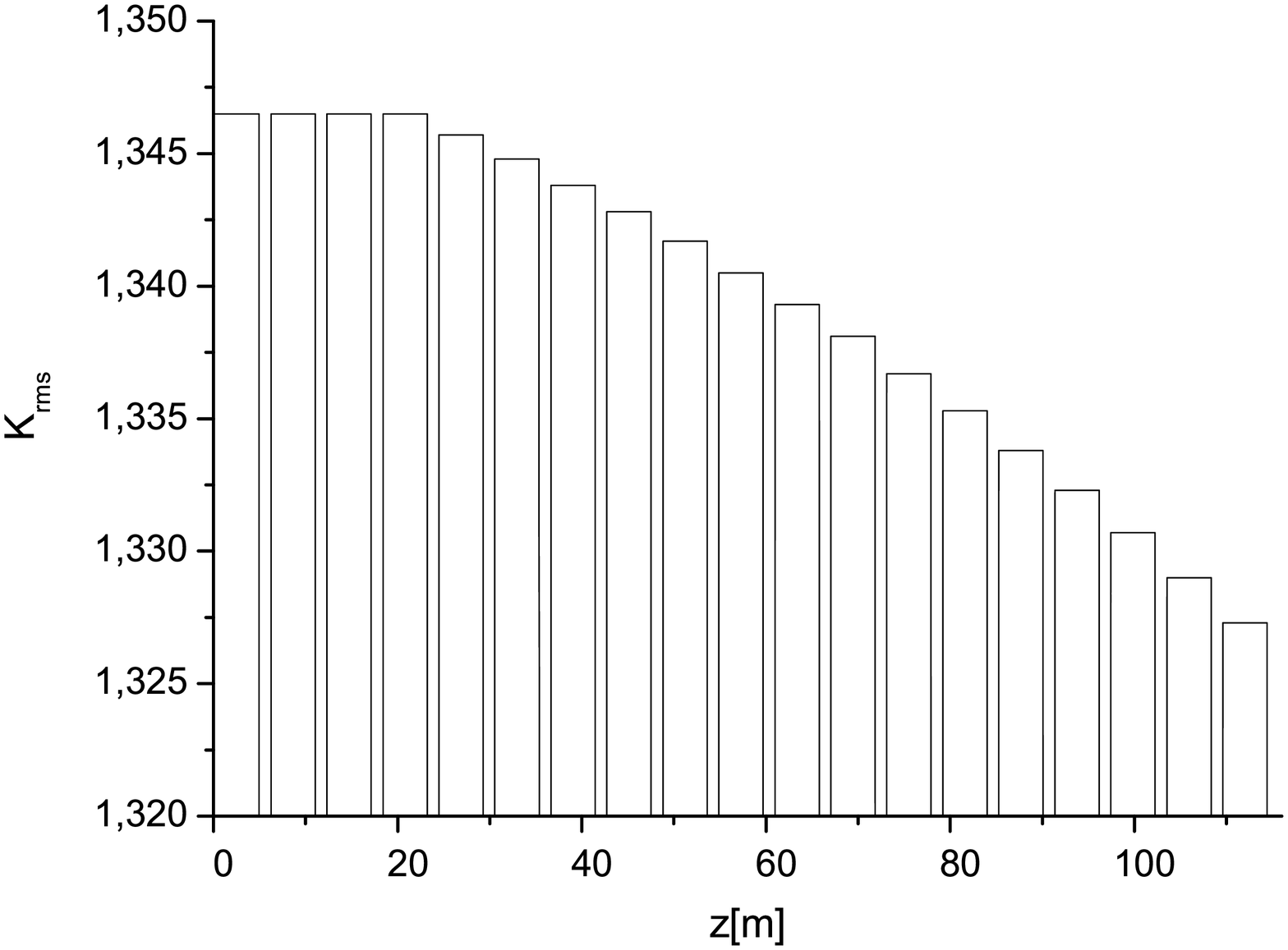}
\caption{Taper configuration for high-power mode of operation at
$0.075$ nm} \label{Tap}
\end{figure}
The tapering law used in the last $19$ cells is shown in Fig.
\ref{Tap}.

\begin{figure}[tb]
\includegraphics[width=0.5\textwidth]{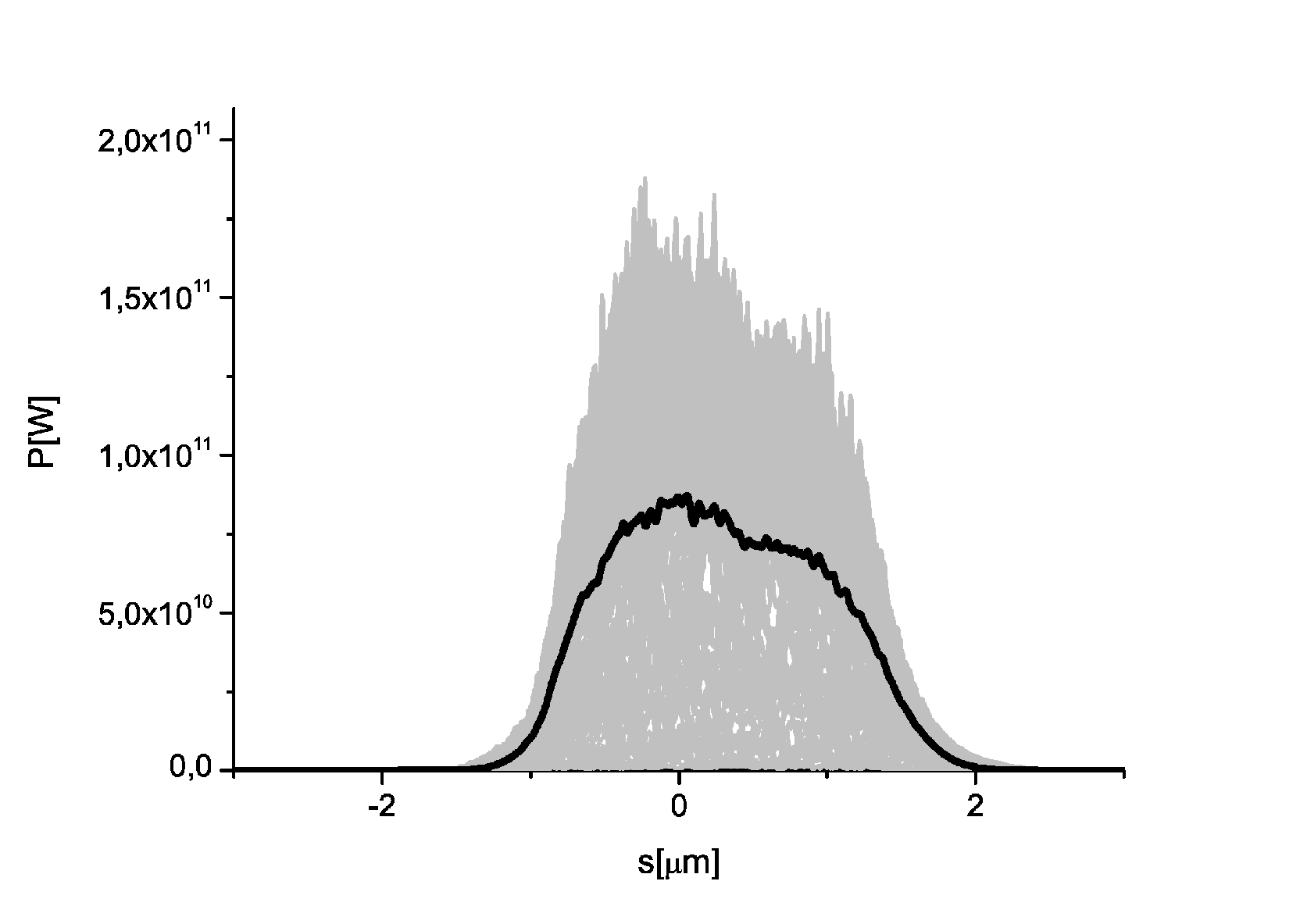}
\includegraphics[width=0.5\textwidth]{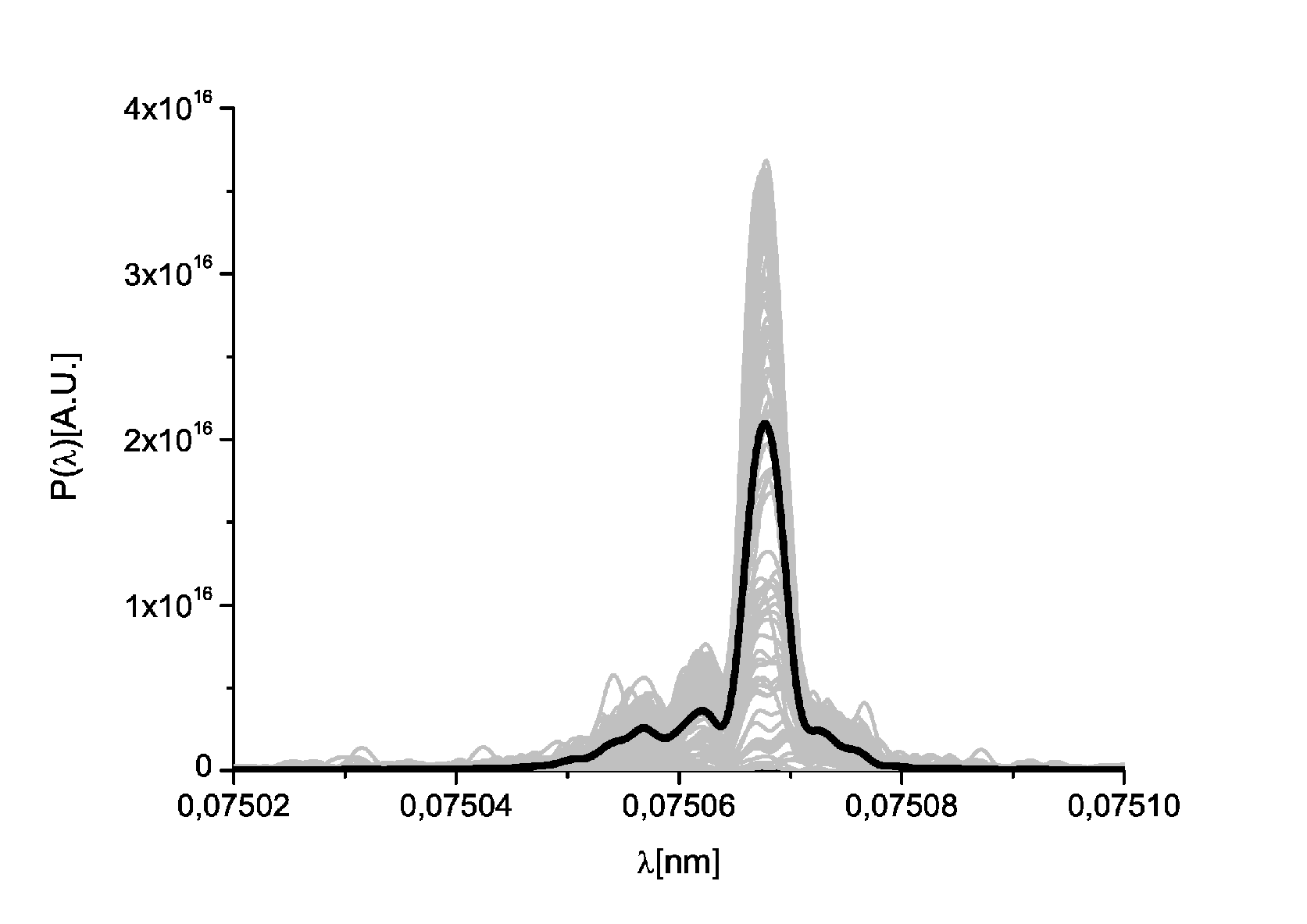}
\caption{(Left plot) Output power and (right plot) output spectrum
after the second part of the output undulator ($19$ cells) tuned at
$0.075$ nm. Grey lines refer to single shot realizations, the black
line refers to the average over a hundred realizations.}
\label{Out2}
\end{figure}
\begin{figure}[tb]
\includegraphics[width=0.5\textwidth]{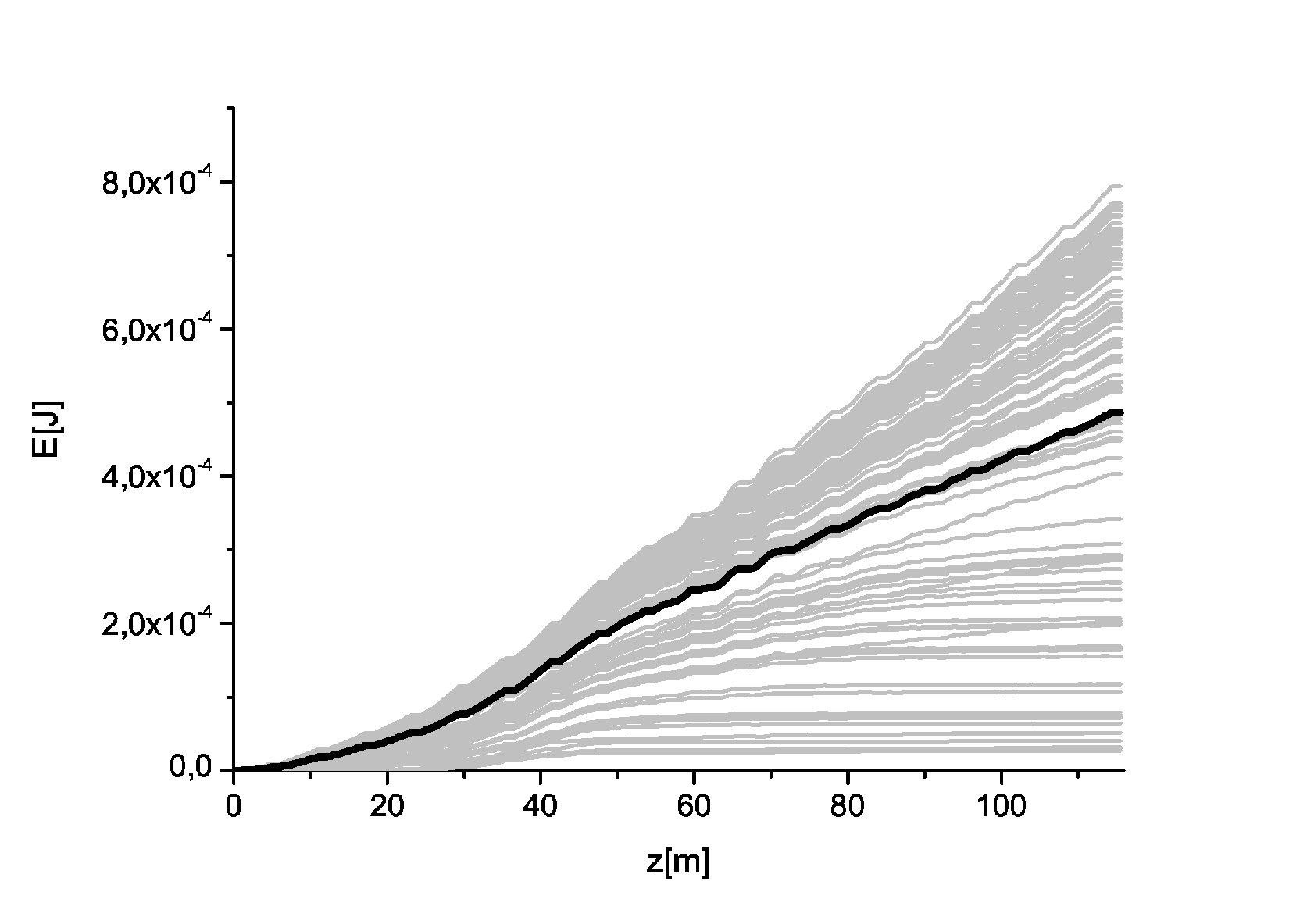}
\includegraphics[width=0.5\textwidth]{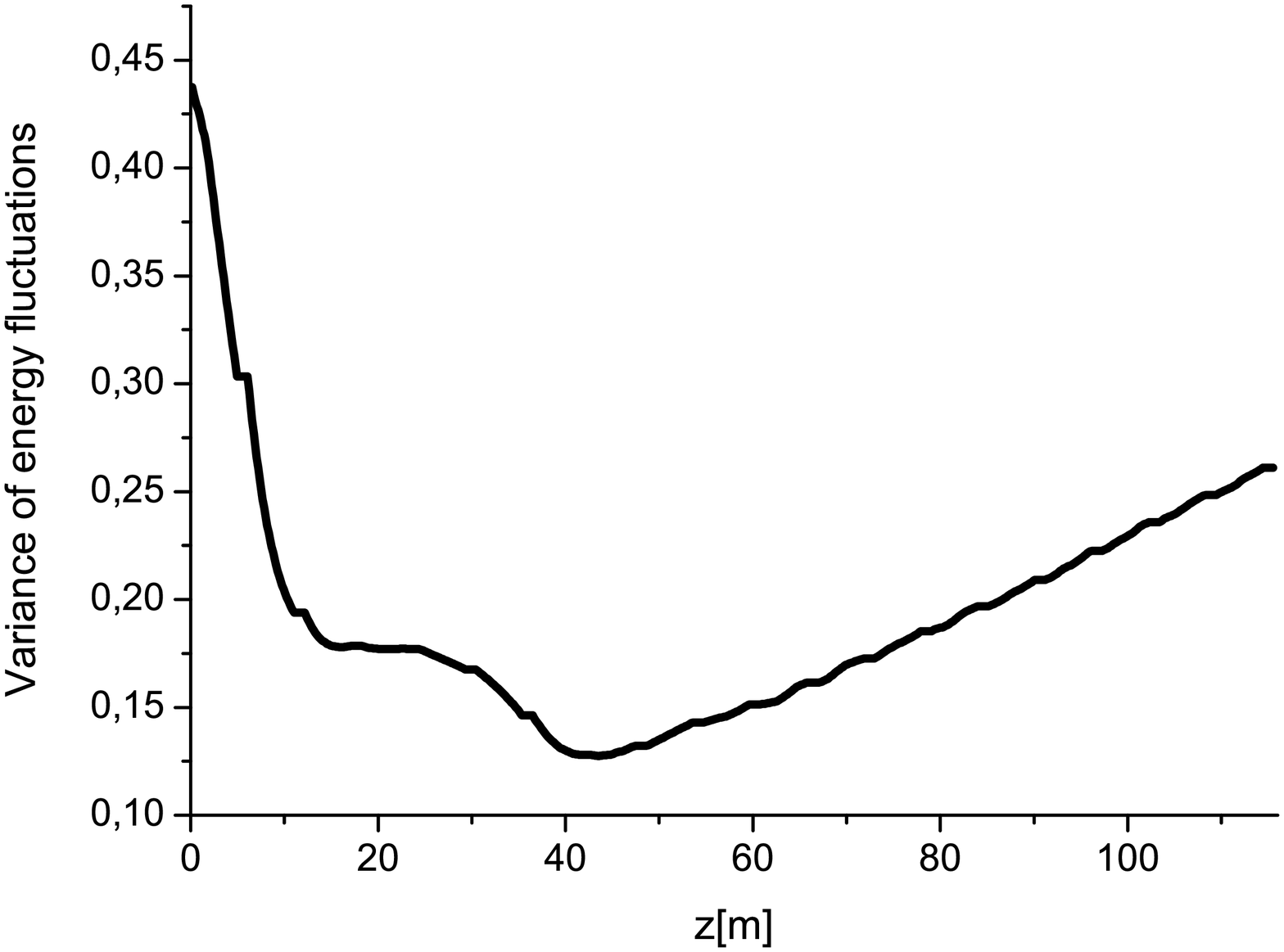}
\caption{(Left plot) Energy in the X-ray radiation and (right plot)
variance of the energy fluctuations of the photon pulse versus the
length of the second part of the output undulator tuned at $0.075$
nm. Grey lines refer to single shot realizations, the black line
refers to the average over a hundred realizations.} \label{Envarout}
\end{figure}
\begin{figure}[tb]
\includegraphics[width=0.5\textwidth]{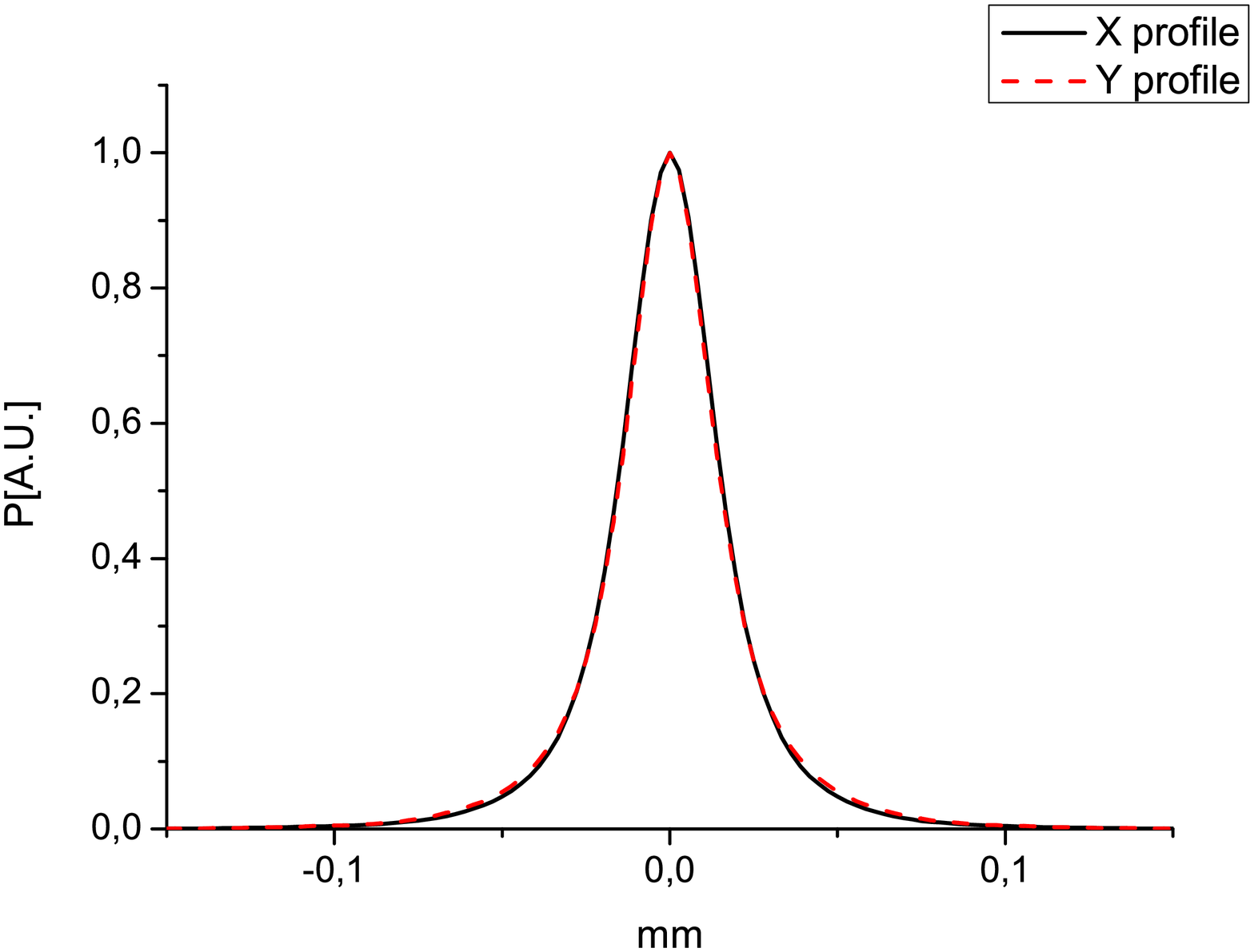}
\includegraphics[width=0.5\textwidth]{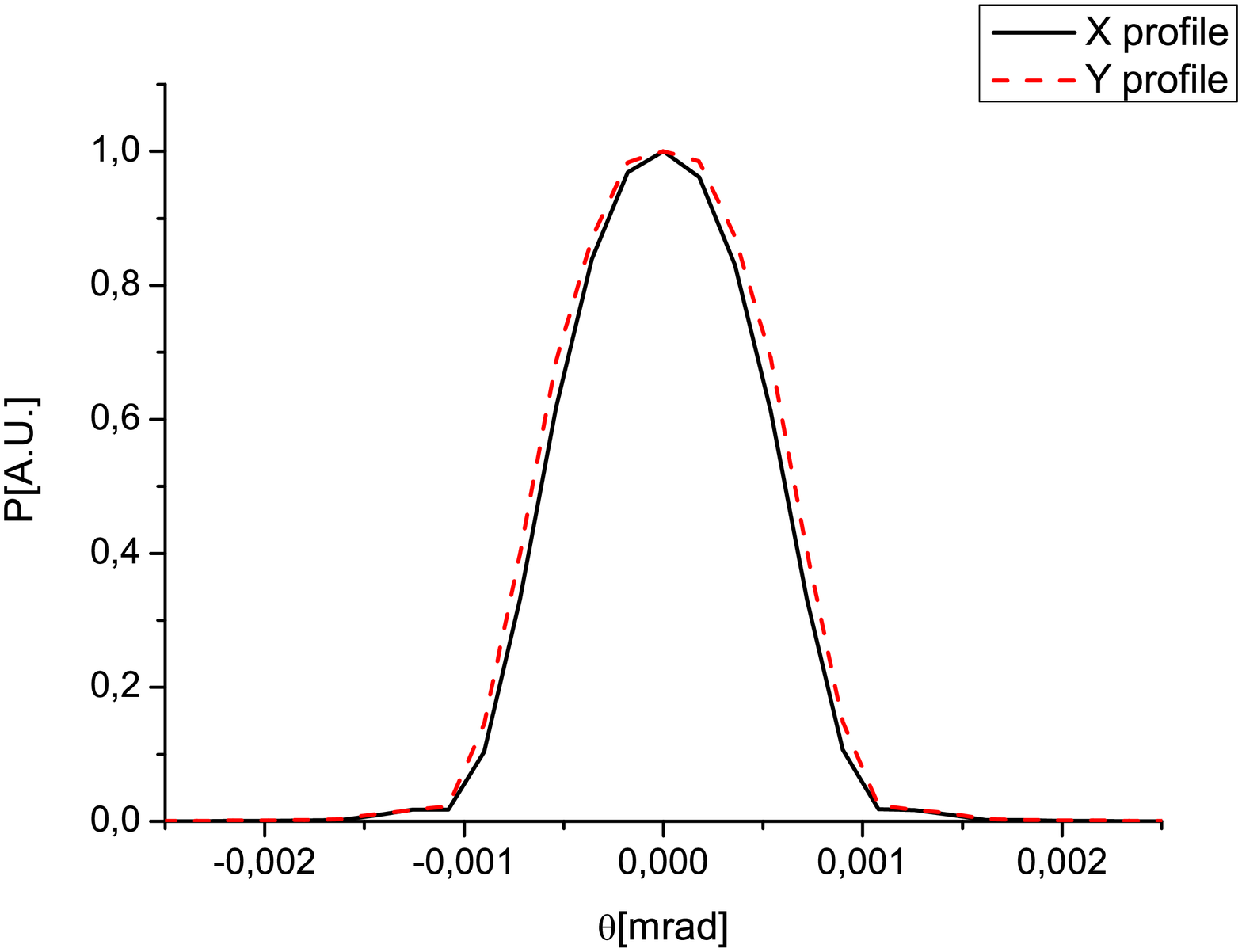}
\caption{(Left plot) Transverse plot of the X-ray radiation pulse
energy distribution and (right plot) angular plot of the X-ray
radiation pulse energy distribution after the second part of the
output undulator tuned at $0.075$ nm.} \label{Enout}
\end{figure}
The output power and spectrum of the entire setup, that is after the
second part of the output undulator ($19$ cells) tuned at $0.075$
nm, is shown in Fig. \ref{Out2}. The evolution of the output energy
and of the variance of the energy fluctuations in the photon pulse
are reported in Fig. \ref{Envarout}. Finally, the expected
transverse size and divergence of the electron beam can be seen in
Fig. \ref{Enout}.

\section{\label{sec:conc} Conclusions}

In this work we addressed the potential for enhancing the
capabilities of the European XFEL. In the hard X-ray regime, a high
longitudinal coherence, in addition to full transverse coherence,
will be the key to a performance upgrade. High longitudinal
coherence is achievable based on a single-crystal self-seeding
scheme, which has been studied for the European XFEL parameters in
\cite{OURY5}. This scheme is compact and can be straightforwardly
installed in the baseline undulator with minimal modifications and
virtually no operational risk. With the radiation beam
monochromatized down to the transform limit, the output power of the
European XFEL could be increased by tapering the tunable-gap
baseline undulator. In particular, in \cite{OURY3} we proposed a
scheme suitable for the European XFEL, to generate TW-level, fully
coherent pulses at photon energies of $8$ keV.

In this paper we proposed a study of the performance of a
self-seeding scheme with single-crystal monochromators for the
European XFEL at X-rays energies higher than $8$ keV. By combining
the two techniques of self-seeding and undulator tapering, we found
that $100$ GW X-ray transform -limited pulses at the photon energy
of $16$ keV can be generated without modification to the TW mode of
operation at photon energy of $8$ keV.

This paper also describes an efficient way for obtaining a many-user
facility, based on the high photon energy and high monochromaticity
of the output radiation expected from our concept. We propose a
photon beam distribution system based on the use of crystals in the
Bragg reflection geometry as deflectors. About $99 \%$ reflectivity
can be achieved for monochromatic X-rays. Angular and bandwidth
acceptances of crystal deflectors match bandwidth and divergence of
the X-ray beam. Therefore, it should be possible to deflect the full
radiation pulse of an angle of order of a radian without
perturbations. The proposed photon beam distribution system would
allow to switch the hard X-ray beam quickly between many instruments
in order to make a more efficient use of the European XFEL source.

\section{Acknowledgements}

We are grateful to Massimo Altarelli, Reinhard Brinkmann,
Serguei Molodtsov and Edgar Weckert for their support and their interest during the compilation of this work.

\end{document}